\documentclass{JHEP3}

\title{Extended set of Majorana spinors, a new dispersion relation,
and a  preferred frame}
\author{D. V. Ahluwalia-Khalilova\thanks{CONACyT (Mexico) is  acknowledged 
for funding this research through Project
32067-E. IUCAA (India) is
is acknowledged for its hospitality, 
and a  senior visiting
professorship which supported part of this work.}\\
Department of Mathematics, 
University of Zacatecas (UAZ), Zacatecas, Zac 98060, Mexico\\
E-mail: \email{d.v.ahluwalia-khalilova@heritage.reduaz.mx}}

\abstract{
We aspire to fufill Majorana's original goal of
bringing full symmetry between the charged and 
fundamentally neutral particles.
We present a description of fundamentally neutral 
particles without any reliance on Dirac spinors.
We show that the extended set Majorana spinors  
(a) describe a  Wigner class of fermions in which 
the charge conjugation and the parity operators
commute, rather than anticommute, and  (b) support two type
of dispersion relations,  the usual $E=\pm\sqrt{p^2+m^2}$, and
 a new $E=- 2m \pm \sqrt{p^2+m^2}$. The latter may be preferred
over the former on minimization of energy requirement
and offers a natural source for observed 
matter-antimatter asymmetry in the universe.
The ensuing physical interpretation requires
existence of a preferred frame which may be identified with the
cosmic neutrino
background.}


\dedicated{Written to celebrate birthday of
Dr. I. S. Ahluwalia-Khalilova}

\begin{document}

\def\beq{\begin{eqnarray}}
\def\eeq{\end{eqnarray}}

\def\ua{\{-,+\}}
\def\da{\{+,-\}}


\def\s{\mbox{\boldmath$\displaystyle\mathbf{\sigma}$}}
\def\J{\mbox{\boldmath$\displaystyle\mathbf{J}$}}
\def\K{\mbox{\boldmath$\displaystyle\mathbf{K}$}}
\def\A{\mbox{\boldmath$\displaystyle\mathbf{A}$}}
\def\B{\mbox{\boldmath$\displaystyle\mathbf{B}$}}

\def\P{\mbox{\boldmath$\displaystyle\mathbf{P}$}}
\def\p{\mbox{\boldmath$\displaystyle\mathbf{p}$}}
\def\hp{\mbox{\boldmath$\displaystyle\mathbf{\widehat{\p}}$}}
\def\x{\mbox{\boldmath$\displaystyle\mathbf{x}$}}
\def\0{\mbox{\boldmath$\displaystyle\mathbf{0}$}}
\def\bv{\mbox{\boldmath$\displaystyle\mathbf{\varphi}$}}
\def\hbv{\mbox{\boldmath$\displaystyle\mathbf{\widehat\varphi}$}}

\def\bn{\mbox{\boldmath$\displaystyle\mathbf{\nabla}$}}

\def\bl{\mbox{\boldmath$\displaystyle\mathbf{\lambda}$}}
\def\bl{\mbox{\boldmath$\displaystyle\mathbf{\lambda}$}}
\def\br{\mbox{\boldmath$\displaystyle\mathbf{\rho}$}}
\def\bfhh{\mbox{\boldmath$\displaystyle\mathbf{(1/2,0)\oplus(0,1/2)}\,\,$}}

\def\mn{\mbox{\boldmath$\displaystyle\mathbf{\nu}$}}
\def\amn{\mbox{\boldmath$\displaystyle\mathbf{\overline{\nu}}$}}

\def\mne{\mbox{\boldmath$\displaystyle\mathbf{\nu_e}$}}
\def\amne{\mbox{\boldmath$\displaystyle\mathbf{\overline{\nu}_e}$}}
\def\rlh{\mbox{\boldmath$\displaystyle\mathbf{\rightleftharpoons}$}}

\def\wm{\mbox{\boldmath$\displaystyle\mathbf{W^-}$}}
\def\hh{\mbox{\boldmath$\displaystyle\mathbf{(1/2,1/2)}$}}
\def\h00h{\mbox{\boldmath$\displaystyle\mathbf{(1/2,0)\oplus(0,1/2)}$}}
\def\znbb{\mbox{\boldmath$\displaystyle\mathbf{0\nu \beta\beta}$}}



\section{Introduction}

Early in December 2001, the long sought-after signal from experiments on neutrinoless
double beta decay was finally reported by the  Heidelberg-Moscow (HM)
collaboration  \cite{HM2001,HM2002}. In its most natural explanation 
the HM events suggest neutrinos to be fundamentally neutral particles 
in the sense of Majorana \cite{EM1937}.\footnote{Parenthetically, we note that
initial concerns of C. E. Aalseth {et al.\/} 
seem to have been fully attended.  
Of the nine original questions
raised by  C. E. Aalseth {et al.\/} six have been withdrawn by the 
authors themselves and the remaining have now been  attended in detail 
by  H. V. Klapdor-Kleingrothaus {\em et al.\/} in Ref. \cite{HM2002}.
The  C. E. Aalseth {et al.\/} ``Comment,'' after extensive revisions,
is now published, see Ref. \cite{cea}.}
 However, fundamental neutrality
of particles is not a neutrino specific property.  For example, in 
supersymmetric theories a host of similarly neutral particles are
required to exist. 
\TABLE{
{\begin{tabular}{|l|l|} \hline\hline
$(1/2,0)\oplus(0,1/2)$ &   $(1/2,0)\oplus(0,1/2)$ \\ \hline\hline
$\{C,P\}=0$ &  $\left[C,P\right]=0$ \\  
{\sc Fermionic  Matter Fields} &  
{\sc Fermionic Gauge Fields}\\ \hline
 $(1,0)\oplus(0,1)$ & $(1/2,1/2)$ \\
$\{C,P\}=0$ & $\left[C,P\right]=0$ \\  
{\sc Bosonic Matter Fields} &  {\sc Bosonic Gauge Fields} \\ \hline\hline
\end{tabular}}
\label{tab1}
\caption{ 
The diagonal Wigner blocks are the ordinary fermionic matter and  
bosonic  gauge fields, while the off-diagonal blocks refer to new
structure in spacetime.}
}

The discovery of fundamentally neutral particle of spin one half
should be considered a major step towards a possible discovery  
of supersymmetry as I noted  at Beyond the Desert
2002, see \cite{bd,A2002}. 
This assertion, in part, arises from our ability to 
construct Table \ref{tab1}. In it
the diagonal Wigner blocks are the ordinary fermionic-matter and 
bosonic-gauge fields. 
The top off-diagonal block is one of the recent key results.
Supersymmetric fermionic gauge particles live in that
block. The bottom off-diagonal block is populated by bosonic matter fields
and awaits experimental confirmation.
It was constructed in a 1993 paper of mine with
Johnson and Goldman cited here as Ref. \cite{AJG1993}. 
The gauge aspect of the  top off-diagonal block
tentatively refers either to the gauginos, or for neutrinos,
should they be confirmed to be Majorana, it should be
interpreted as an internal fermionic line as it appears  
in neutrinoless double beta decay. 
The $C$ and $P$ operators belong to the indicated representation space for 
each of the Wigner blocks. Possibility, but without explicit construction 
(for which we take credit), of such blocks is due to Wigner, and his 
colleagues \cite{EPW1962}.

The existing description of Majorana particles has a pivotal reliance
on Dirac spinors \cite{EM1937} \textendash~ a circumstance I, and many 
other scholars who concern themselves with such questions, 
consider unsatisfactory.  The competing 
reformulation of McLennan and Case \cite{Mc1957,Case1957}, which 
avoids its dependence on Dirac spinors, is on the other hand 
incomplete. It only carries two degrees freedom to span a four 
dimensional $(1/2,0)\oplus(0,1/2)$ representation space and cannot
address a whole range of relevant questions.

For sometime, therefore, we have been attempting to understand 
the type of particles Majorana envisaged \cite{AGJ1994,DVA1996}. 
This paper is the first
exposition in the series which we think places Majorana's vision 
fully at par with that of Dirac on charged particles.

\section{Review of Dirac Construct}

To better appreciate the construct that we present let us first briefly
review the structure of Dirac's construct. This will also help us set up 
the notation.
A Dirac spinor, in Weyl representation,
is
\beq
\psi(\p) = \left(
\begin{array}{c}
\phi_R(\p) \\
\phi_L(\p)
\end{array}
\right)
\eeq 
where the massive  Weyl  spinors $\phi_R(\p)$ transforms as
$(1/2,0)$ representation-space objects, and  massive  Weyl  spinors $\phi_L(\p)$
transforms as  $(0,1/2)$ representation-space objects. 
Our notation is very close to that of Ryder \cite{LHR1996}.
For the ease of reference we shall mark needed observations with ${\cal{O}}_n$, with
$n=1,2,3,\ldots$. 

${\cal{O}}_1$: 
The first thing the reader should explicitly note is that
the Dirac spinors are constructed
by taking  both   $\phi_R(\p)$ and  $\phi_L(\p)$ to be
in same helicity. Contrary is required for Majorana spinors (see below).

Following the usual particle-antiparticle nomenclature,
the following holds in addition:

${\cal{O}}_2$: 
In the rest frame, characterized by $\p=\0$, for
particles
there is no relative phase  between $\phi_R(\p)$ and  $\phi_L(\p)$,
i.e., $\phi_R(\0)= \phi_L(\0)$. Whereas, for
antiparticles
  $\phi_R(\0)$ and  $\phi_L(\0)$ carry opposite phase
i.e., $\phi_R(\0)= - \phi_L(\0)$.

\def\rb{\kappa^{\left(\frac{1}{2},0\right)}}
\def\lb{\kappa^{\left(0,\frac{1}{2}\right)}}

This last property has only been 
noted explicitly in the last decade.\footnote{See, my book review 
of Ryder's text on quantum field theory in Ref. \cite{review}, and 
 Gaioli and  Garcia-Alvarez's American Journal of Physics disposition
in Ref. \cite{GGA1995}.  
The genesis of this observation in fact begins with 
Ref. \cite{AJG1993} formally, and a conversation between 
myself and Christoph Burgard at Texas A\&M University when we both
were there as students.}
The momentum-space wave equation satisfied by the spinors thus
constructed follows uniquely from 
\cite{LHR1996,AK2001},\footnote{Note \cite{AJG1993,GGA1995,review}, 
the necessity of minus sign in Eq. (\ref{p44}). This sign in ``group
theoretical derivations'' of Dirac equation has been consistently missed 
(see, e.g., \cite{LHR1996} p.44 in 1987 edition, and pp. 168-169 of
Ref. \cite{JH1999}.
The minus is necessary to have antiparticles in the momentum space. The 
antiparticles, in the spacetime description with minus sign dropped, appear
only after ``two mistakes which cancel one mistake of 
omitting the indicated sign.''}

\beq
\phi_R (\0) = \pm \phi_L(\0) \,, \label{p44}
\eeq
and
\beq
\phi_R(\p) = \rb \phi_R (\0)\,, 
\quad
\phi_L(\p) = \lb \phi_L (\0)\,.
\eeq 
Here, 
\beq
\kappa^{\left(\frac{1}{2},0\right)} &=&\exp\left(+\, 
\frac{\s}{2}\cdot\bv\right)= \sqrt{\frac{E+m}{2\,m}}
\left(\mathbb{I}+\frac{\s\cdot\p}{E+m}\right)
\,,\label{br}\\
\kappa^{\left(0,\frac{1}{2}\right)}&=&\exp\left(-\, 
\frac{\s}{2}\cdot\bv\right)= \sqrt{\frac{E+m}{2\,m}}
\left(\mathbb{I}-\frac{\s\cdot\p}{E+m}\right)\,,\label{bl}
\eeq
with the  boost parameter  defined as:
\beq
\cosh(\varphi)=\frac{E}{m},\quad
\sinh(\varphi)=\frac{\vert\p\vert}{m},\quad 
{\hbv}
=\frac{\p}{\vert \p\vert}\,.\label{bp}
\eeq
These  wave equations are,
\beq
\left(\gamma^\mu p_\mu \mp m \mathbb{I}\right)\psi(\p)=0\,.\label{zzz}
\eeq
Here, $\mathbb{I}$ are $n\times n$ identity matrices. 
Their dimensionality being apparent from the context in which
they appear.\footnote{So, e.g., 
in Eqs. (\ref{br}) and (\ref{bl}), the
 $\mathbb{I}$ stand for $2\times 2$ identity matrices; while
in Eq. (\ref{zzz})  $\mathbb{I}$  is a $4\times 4$ identity matrix.}   
The $\gamma^\mu$ have
their standard Weyl-representation form:
\beq
\gamma^0= \left(
\begin{array}{cc}
\mathbb{O} & \mathbb{I} \\
\mathbb{I} & \mathbb{O}
\end{array}\right)
\,\quad 
\gamma^i=
\left(
\begin{array}{cc} \mathbb{O} & - \sigma^i \\
\sigma^i & \mathbb{O}\end{array}
\right)\,.\label{gammamatrices}
\eeq
For consistency of notation, $\mathbb{O}$ 
here represents a $n\times n$ 
null matrix (in the above equation, $n=2$).
Letting,
$
p_\mu = i\hbar \partial_\mu,
$
and, 
$
\psi(x) = \exp\left( \mp \frac{i}{\hbar} p_\mu x^\mu\right)\psi(\p),
$
with upper sign for particles, and lower sign for antiparticles,
one obtains the configuration space Dirac equation:
$
\left(i \gamma^\mu \partial_\mu -  m \mathbb{I}\right)\psi(x)=0.
$

${\cal{O}}_3$: 
It should be noted that in Eq. (\ref{bp}), $E > 0$. Whereas,
$Det \left(\gamma^\mu p_\mu \mp m \mathbb{I}\right) = 0$ yields
$E = \pm \sqrt{m^2+p^2}$. At this stage, going back to original
paper of Dirac, Dirac did not consider this as an internal 
inconsistency and did not discard the  $E < 0 $ through 
a constraint. He could have done so in a covariant manner
and hardly any one would raised an objection. The lesson is simple:
one should not impose mathematical constraints to satisfy one's
physical intuition.\footnote{
The physical intuition may ask for $E>0$, or a definite spin for
particles, etc. Such constraints may have limited validity in
a classical framework. But in a quantum framework 
interactions shall, in general, induce transitions between classically allowed
and classically forbidden sectors. }
 Had Dirac taken the path of physical intuition,
rather than  opting for a mathematically imposed inevitability, a local
$U(1)$ gauge theory based on such a (covariant)theory 
would have been mathematically pathological, and 
physically it would have missed Lamb shift, not to say antiparticles.  
We have tried to teach the same lesson in a different context in 
Ref. \cite{KA2002,AK2001}
to the exasperation of many distinguished people. The same lesson shall
be seen to be important for Majorana particles now: a four dimensional
$(1/2,0)\oplus(0,1/2)$ representation space  \textemdash~ whether it be
Dirac, or Majorana \textemdash~ requires four degrees of freedom (i.e.,
four independent spinors).\footnote{That these {\em four\/} degrees of freedom
may be complex, or real (in a given realization), is a different matter.
We shall not contest if one wishes to say that in ``Majorana representation 
(realization)'' 
Dirac spinors has eight real degrees of freedom,
and Majorana has four.}

The derivation of Dirac  equation as outlined here 
carries a quantum mechanical aspect
in allowing for the fact the the two Weyl spaces may carry a relative
phase, in the sense made explicit above; and concurrently a relativistic
element via the Lorentz transformation properties of the Weyl spinors.
In turn the very existence of the latter depends on existence of
left and right spacetime $SU(2)$s:
\beq
SU(2)_R:\quad \A= \frac{1}{2}\left(\J+ i\K\right)\,,\quad
SU(2)_L:\quad \B= \frac{1}{2}\left(\J- i\K\right)\,.
\eeq
The $\J$ and $\K$ represent the generators of rotations and boosts
for the any of the relevant finite dimensional 
representation space which may be under consideration.
From the womb of this structure emerges a new symmetry, i.e., that
of charge conjugation. The operator associated with this symmetry shall
be now written in a slightly unfamiliar form (so as to fully exploit 
it for understanding Majorana particles):
\beq
{C} = 
\left(
\begin{array}{cc}
\mathbb{O} & i\,\Theta \\
-i\,\Theta & \mathbb{O}
\end{array}
\right) {K}\,.\label{cc}
\eeq
Here, operator $K$ complex conjugates any Weyl spinor that appears 
on its right, 
and  $\Theta$ is the Wigner's spin-$1/2$ time reversal 
operator\footnote{
For an arbitrary spin it is defined 
by the property $\Theta \J \Theta^{-1}= -\,\J^\ast$.
We refrain from  identifying $\Theta$ with ``$-\,i\, \sigma_2$,''
as is done implicitly in all considerations
on the subject -- see, e.g., Ref. \cite{PR1989} --  because
such  an identification does not exist for higher-spin $(j,0)\oplus(0,j)$ 
representation spaces. The existence of  Wigner time reversal
operator for all $j$, allows, for fermionic $j$'s,
the introduction of $(j,0)\oplus (0,j)$ neutral particle spinors.}
\beq
\Theta=
\left(
\begin{array}{cc}
0 & -1 \\
1 & 0
\end{array}
\right)\,.\label{wt}
\eeq
It is readily seen that the standard form, $C=-\gamma^2 K$, is recovered.
It is important to note, in the context
of the derivation of a wave equation for the extended set 
of Majorana spinors (presented below), that $(1/2,0)\oplus(0,1/2)$ boost
operator, $\rb\oplus \lb$, and the 
$(1/2,0)\oplus(0,1/2)$-space charge conjugation operator, $C$, commute.

So particles and antiparticles are offsprings of a fine interplay between
the quantum realm and the realm of spacetime symmetries.
Here, we have made it transparent.
The operation of C takes, up to a spinor-dependent global phase, 
the particle spinors into antiparticle spinors
and vice versa -- see, Eq. (\ref{ceq}) of Appendix G .
Keeping with our pedagogic style, we note:
The Dirac spinors are thus not eigenspinors of the Charge conjugation
operator. That honor belongs to extended set of Majorana 
spinors which we introduce below.

Since our task is a logical one, rather than a historical one, we now
ask what are the eigenspinors of the C operator?
The answer which we shall obtain is: a set of four spinors; two of which 
are identical to massive McLennan-Case spinors (and are 
called Majorana spinors in literature), and other two which
are new. The set of four spinors representing fundamentally neutral
spinors shall be called {\em extended set of Majorana spinors.\/}
Towards the task of obtaining 
the extended set of Majorana spinors we shall
follow a path which sheds maximal light on the various relative phases
involved.  At the same time our procedure  takes due note of the
the transformation properties of the right and left transforming 
components of these spinors and makes them manifest while at the
same time emphasizing their helicity orientations. 
One may look at our defined task as to first build counterpart 
of Dirac's $\psi(\p)$ and then to fully outline the properties 
of this counterpart in a way that makes it stand 
in its own right.

\section{Extended set of Majorana spinor: Eigenspinors of charge 
conjugation operator }

In the spirit just outlined our task begins with the observation 
that
\beq
\left( \kappa^{\left(0,\frac{1}{2}\right)} \right)^{-1} = 
\left( \kappa^{\left(\frac{1}{2},0\right)} \right)^\dagger\,,\quad
\left( \kappa^{\left(\frac{1}{2},0\right)} \right)^{-1} = 
\left( \kappa^{\left(0,\frac{1}{2}\right)} \right)^\dagger \, .
\eeq
Further,   $\Theta$, the Wigner's spin-$1/2$ time reversal 
operator, has the property
\beq
\Theta \left[\s/2\right] \Theta^{-1} = -\, \left[\s/2\right]^\ast\,,  
\label{wigner}
\eeq
When combined, these observations imply that \cite{PR1989}: 
(a)
If $\phi_L(\p)$ transforms as a left handed spinor, then
$\left(\zeta_\lambda \Theta\right) \,\phi_L^\ast(\p)$
transforms as a right handed spinor -- where, $\zeta_\lambda$ is
an unspecified phase;
(b)
If $\phi_R(\p)$ transforms as a right handed spinor, then
$\left(\zeta_\rho \Theta\right)^\ast \,\phi_R^\ast(\p)$
transforms as a left handed spinor -- where, $\zeta_\rho$ is
an unspecified phase.
As a consequence, the following spinors 
belong to the $(1/2,0)\oplus(0,1/2)$
representation space :

\beq
\lambda(\p) =
\left(
\begin{array}{c}
\left(\zeta_\lambda \Theta\right) \,\phi_L^\ast(\p)\\
\phi_L(\p)
\end{array}
\right)\,,\quad
\rho(\p)=
\left(
\begin{array}{c}
\phi_R(\p)\\
\left(\zeta_\rho \Theta\right)^\ast \,\phi_R^\ast(\p)
\end{array}
\right)\,.
\eeq
Demanding $\lambda(\p )$ and $\rho (\p )$ to 
be self/anti-self conjugate under  $C$, 
\beq
{C} \lambda(\p) = \pm  \lambda(\p)\,,\quad
 {C} \rho(\p) = \pm  \rho(\p)\,,
\eeq
restricts the phases, $\zeta_\lambda$ and
$\zeta_\rho$, to two values:
\beq
\zeta_\lambda= \pm\,i\,,\quad \zeta_\rho=\pm\,i\,.
\eeq
The plus sign in the above equation
yields self conjugate, $\lambda^S(\p)$ and $\rho^S(\p)$ 
spinors; while the minus
sign results in the anti-self conjugate spinors,  $\lambda^A(\p)$ and
$\rho^A(\p)$.\footnote{ 
Since,
$
i\Theta = \sigma_2
$,
we may write:
\beq
\lambda(\p) =
\left(
\begin{array}{c}
\pm \sigma_2 \,\phi_L^\ast(\p)\\
\phi_L(\p)
\end{array}
\right)\,,\quad
\rho(\p)=
\left(
\begin{array}{c}
\phi_R(\p)\\
\mp\sigma_2 \,\phi_R^\ast(\p)
\end{array}
\right)\,.
\eeq
where the upper sign is for self conjugate spinors, and the lower
sign yields the antiself conjugate spinors.
The $\lambda^S(\p)$ thus turn out to be identical to the standard
textbook, or McLennan-Case, Majorana spinors. $\lambda^A(\p)$,
are new and mathematically orthogonal to $\lambda^S(\p)$. The
use, and linear dependence of the $\rho(\p)$ on $\lambda(\p)$,
shall be discussed as we proceed further.} 

\medskip
In the rest frame, it is clear that the phase relationship
between the right handed and left handed Weyl components
of the fundamentally neutral spinors
is profoundly different from their Dirac conunterparts.
Furthermore, as shall be explicitly shown immediately,
the two Weyl components of a   fundamentally neutral spinor
must carry {\em opposite\/} helicities. That is, Majorana spinors
cannot be eigenspinors of the helicity operator. This, we
believe, is important for making a physical picture
of neutrinoless double beta decay in which a neutrino is 
emitted as a particle at one vertex 
and absorbed as antiparticle at another vertex. In other words,
the fact that
fundamentally neutral spinors are not single helicity objects,
but invite an interpretation of dual helicity spinors, makes
processes such as neutrinoless double beta decay possible.

To obtain explicit expressions for $\lambda (\p )$, we first write down 
the rest spinors. These are:
\beq
\lambda^S(\0) = 
\left(
\begin{array}{c}
+\,i \,\Theta \,\phi_L^\ast(\0)\\
\phi_L(\0)
\end{array}
\right)\,,\quad
\lambda^A(\0) = 
\left(
\begin{array}{c}
-\,i \,\Theta \,\phi_L^\ast(\0)\\
\phi_L(\0)\, 
\end{array}
\right)\, .
\eeq
Next, we choose the $\phi_L(\0)$ to be helicity eigenstates,
\beq
\s\cdot{\hp} \;\phi_L^\pm (\0)= \pm\;\phi_L^\pm(\0)\,,
\label{x}
\eeq  
and concurrently note that\footnote{
See Appendix A for derivation of Eq. (\ref{y}).
The explicit forms of $\phi^\pm_L(\0)$ 
are given in Appendix B.}
\beq
\s\cdot\hp \, \Theta \left[\phi_L^\pm (\0)\right]^\ast
= \mp\, \Theta\left[\phi_L^\pm(\0)\right]^\ast\,.
\label{y}
\eeq 
That is, $\Theta \left[\phi_L^\pm (\0)\right]^\ast$
has opposite helicity of $\phi_L^\pm (\0)$.
Since $\s\cdot\hp$ commutes with the boost operator 
$\kappa^{\left(1/2,0\right)}$ the above result applies for all
momenta. In conjunction with the definition of the 
neutral spinors we are thus lead to the result that
neutral spinors are {\em not\/} single helicity objects. Instead,
they invite an interpretation of dual helicity spinors.
In the process  we are led to four rest spinors. 
Two of which are self-conjugate, 
\beq
\lambda_{\{-,+\}}^S(\0) = 
\left(
\begin{array}{c}
+\,i \,\Theta \,\left[\phi^+_L(\0)\right]^\ast\\
\phi^+_L(\0)
\end{array}
\right)\,,\quad
\lambda_{\{+,-\}}^S(\0) = 
\left(
\begin{array}{c}
+\,i \,\Theta \,\left[\phi^-_L(\0)\right]^\ast\\
\phi^-_L(\0)\, 
\end{array}
\right)\, , \quad \label{ls}
\eeq
and the other two, which are anti-self conjugate,
\beq
\lambda_{\{-,+\}}^A(\0) = 
\left(
\begin{array}{c}
-\,i \,\Theta \,\left[\phi^+_L(\0)\right]^\ast\\
\phi^+_L(\0)
\end{array}
\right)\,,\quad
\lambda_{\{+,-\}}^A(\0) = 
\left(
\begin{array}{c}
-\,i \,\Theta \,\left[\phi^-_L(\0)\right]^\ast\\
\phi^-_L(\0)\, 
\end{array}
\right)\, .\label{la}
\eeq
The first helicity entry refers to the $(1/2,0)$ transforming component of the
 $\lambda(\p)$, while the second entry encodes the helicity of
the $(0,1/2)$  component. 
The boosted spinors are now obtained via the operation:
\beq
\lambda_{\{h,-h\}}(\p)=\left(
\begin{array}{cc}
\kappa^{\left(\frac{1}{2},0\right)} & \mathbb{O} \\
\mathbb{O} & \kappa^{\left(0,\frac{1}{2}\right)}
\end{array}
\right)\lambda_{\{h,-h\}}(\0)\,.\label{z}
\eeq
In the boosts, we replace $\s\cdot\p$ by $\s\cdot{\hp}\,\vert \p\vert$,
and then exploit Eq. (\ref{y}). After simplification,
Eq. (\ref{z}) yields:
\beq
\lambda_{\{-,+\}}^S(\p)
=
\sqrt{\frac{E+m}{2\,m}}\left(1-\frac{ \vert \p \vert}{E+m}\right)
\lambda_{\{-,+\}}^S(\0)\,,\label{lsup}
\eeq
which, in the massless limit, {\em identically vanishes;\/} while
\beq
\lambda_{\{+,-\}}^S(\p)
=
\sqrt{\frac{E+m}{2\,m}}\left(1+\frac{ \vert \p \vert}{E+m}\right)
\lambda_{\{+,-\}}^S(\0)\, ,
\label{lsdown}
\eeq
does not.
We hasten to warn the reader that one should not be tempted to read the
two different prefactors to $\lambda^S(\0)$
in the above expressions as the boost operator that appears in Eq. 
(\ref{z}). For one thing, there is only one (not two) 
boost operator(s) in the
$(1/2,0)\oplus(0,1/2)$ representation space. The simplification that
appears here is due to a fine interplay between Eq. (\ref{y}), the boost
operator, and the structure of the $\lambda^S(\0)$.
Similarly, the anti-self conjugate set of the boosted spinors reads:
\beq
\lambda_{\{-,+\}}^A(\p)
=
\sqrt{\frac{E+m}{2\,m}}\left(1-\frac{ \vert \p \vert}{E+m}\right)
\lambda_{\{-,+\}}^A(\0)\,,\label{laup}\\
\lambda_{\{+,-\}}^A(\p)
=
\sqrt{\frac{E+m}{2\,m}}\left(1+\frac{ \vert \p \vert}{E+m}\right)
\lambda_{\{+,-\}}^A(\0)\,.\label{ladown}
\eeq
In the massless limit, the first of these spinors 
{\em identically vanishes\/}, while the second does not.

\section{Majorana Dual}

For any $(1/2,0)\oplus(0,1/2)$ spinor  $\xi(\p)$, whether it be
Majorana or Dirac,
 the 
Dirac dual spinor $\overline{\xi}(\p)$ is defined as:
\beq
\overline{\xi}(\p) = \xi^\dagger(\p) \gamma^0
\eeq 
With the Dirac dual, the Majorana spinors have a imaginary definite 
bi-orthogonal 
norm (see Appendix C). Recalling the  implicitly-contained lessons
in quantization of the Dirac field  we wish to introduce
a dual  which is appropriate for the extended set
of  Majorana spinors. The new dual must have the 
property that it yields an invariant real definite norm. 
In addition, the new dual 
must secure a  positive definite norm for two of
the for spinors contained in the extended set of Majorana spinors, and negative
definite norm
for the remaining two. A unique, up to a relative sign, definition of such
a dual, which we call 
{\em Majorana dual,\/}
for each of the spinors is:
\beq
&& \lambda^S(\p): \quad  \stackrel{\neg}\lambda^S(\p)
= + \left[\rho^A(\p)\right]^\dagger \gamma^0 \, \\
&&  \lambda^A(\p):  \quad  \stackrel{\neg}\lambda^A(\p)= 
- \left[\rho^S(\p)\right]^\dagger \gamma^0 \,, 
\eeq
where the $\rho(\p)$ are given in Appendix D. 
With Majorana dual so defined, we  then have,
\beq
\stackrel{\neg}\lambda^S_{\alpha}(\p)\;
\lambda^S_{\alpha^\prime}(\p) = +\; 2 m\; \delta_{\alpha\alpha^\prime}\,,
\label{zd1}\\
\stackrel{\neg}\lambda^A_{\alpha}(\p)\;
\lambda^A_{\alpha^\prime}(\p) = -\; 2 m \;\delta_{\alpha\alpha^\prime}\,.
\label{z5}
\eeq
The completeness relation,
\beq
\frac{1}{2 m}\sum_\alpha 
 \left[\lambda^S_{\alpha}(\p) \stackrel{\neg}\lambda^S_{\alpha}(\p) 
      - \lambda^A_{\alpha} (\p)\stackrel{\neg}\lambda^A_{\alpha}(\p)\right]
 = \mathbb{I}\,, \label{z1}
\eeq
clearly shows the non-trivial mathematical necessity of the anti-self conjugate
spinors (cf. observation ${\cal{O}}_3$). Equations (\ref{zd1}), (\ref{z5}), and
(\ref{z1}) have their direct counterpart in Dirac's construct -- see,
Eqs. (\ref{d1}), (\ref{5}), and
(\ref{1}) in Appendix F.

\section{A Master wave equation for spinors}

Appendix F shows that extended set of Majorana spinors do not satisfy
Dirac equation. 
Therefore, to study time evolution of neutral particle spinors we need 
appropriate wave equation. This we do in the following manner.
First, we obtain the momentum-space wave equation satisfied
by the $\lambda(\p)$ spinors. The time evolution then follows by careful,
but simple,
implementation of the
``$p_\mu \rightarrow i\partial_\mu$'' prescription.

We seek a momentum-space wave equation
for a general $(1/2,0)\oplus(0,1/2)$ spinor\footnote{Since the method 
we use, and the results we obtain, appear somewhat
unusual we exercise extra care in presenting our derivation. 
We, therefore, present a unified method which applies not only 
to the extended set of Majorana spinors but it applies equally well 
to other cases (such as the Dirac formalism). 
The method is a generalization of the textbook 
procedure \cite{LHR1996} with corrections noted in Refs. 
\cite{AJG1993,GGA1995,review,AK2001}.  }
\beq
\xi(\p) = \left(
\begin{array}{c}
\chi^{\left(\frac{1}{2},0\right)}(\p)\\
 \chi^{\left(0,\frac{1}{2}\right)}(\p)
\end{array}
\right)\,.\label{eq1}
\eeq
In particle's rest frame, where, $\p=\0$, by definition,
\beq
\chi^{\left(\frac{1}{2},0\right)}(\0)
={\mathcal  A}\;
 \chi^{\left(0,\frac{1}{2}\right)}(\0)\,. \label{eq2}
\eeq
Here, the $2\times 2$ matrix ${\mathcal A}$ 
encodes $C$, $P$, and $T$ properties of the spinor
and is left unspecified at the moment except that we require it to
be invertible.
We envisage the most general form of 
$
{\mathcal A}
$ 
to be a unitary matrix with determinant $\pm 1$:
\beq
{\mathcal A_\pm} = 
\left(
\begin{array}{cc}
a\, e^{i \phi_a} & \sqrt{\pm 1-a^2}\, e^{i\phi_b} \\
 - \sqrt{\pm 1-a^2}\, e^{- i\phi_b} & a\,  e^{- i \phi_a} 
\end{array}
\right)\,,\label{amat}
\eeq
with $a$, $\phi_a$, and $\phi_b$ real. The plus sign yields
Determinant of ${\mathcal A}$
to be $+1$, while the minus sign yields it to be $- 1$.
Once 
$\chi^{\left(\frac{1}{2},0\right)}(\0)$ and $
\chi^{\left(0,\frac{1}{2}\right)}(\0)$ are specified
the $\chi^{\left(\frac{1}{2},0\right)}(\p)$ and $
\chi^{\left(0,\frac{1}{2}\right)}(\p)$ follow from,
\beq
\chi^{\left(\frac{1}{2},0\right)}(\p) =&&\rb\;
\chi^{\left(\frac{1}{2},0\right)}(\0) \,, \label{eq3}\\
\chi^{\left(0,\frac{1}{2}\right)}(\p) =&&\lb\;
\chi^{\left(0,\frac{1}{2}\right)}(\0)\,. \label{eq4}
\eeq
Below, we shall need their inverted forms also. These we write
as follows:
\beq
&&\chi^{\left(\frac{1}{2},0\right)}(\0) =
\left(\rb\right)^{-1} \;\chi^{\left(\frac{1}{2},0\right)}(\p)\,,\label{eq3b}\\
&&  \chi^{\left(0,\frac{1}{2}\right)}(\0) =
\left(\lb\right)^{-1}\;\chi^{\left(0,\frac{1}{2}\right)}(\p)\,.\label{eq4b}
\eeq
Equation (\ref{eq2}) implies,
\beq
\chi^{\left(0,\frac{1}{2}\right)}(\0) ={\mathcal A}^{-1}
\chi^{\left(\frac{1}{2},0\right)}(\0)
\eeq
which on immediate use of (\ref{eq3b}) yields, 
\beq
\chi^{\left(0,\frac{1}{2}\right)}(\0) &=&
{\mathcal A}^{-1} \left(\rb\right)^{-1} 
\;\chi^{\left(\frac{1}{2},0\right)}(\p) \,.
\eeq
However, since 
\beq
\left(\rb\right)^{-1} = \lb
\eeq
we have:
\beq
\chi^{\left(0,\frac{1}{2}\right)}(\0) =
{\mathcal A}^{-1}\; \lb \;\chi^{\left(\frac{1}{2},0\right)}(\p)
\,.\label{eq79}
\eeq
Similarly,
\beq
\chi^{\left(\frac{1}{2},0\right)}(\0) =
{\mathcal A}\;\rb \;\chi^{\left(0,\frac{1}{2}\right)}(\p)
\,.\label{eq80}
\eeq
Substituting for  $\chi^{\left(\frac{1}{2},0\right)}(\0)$
from Eq. (\ref{eq80}) in
Eq. (\ref{eq3}) and re-arranging
gives:
\beq
-\;\chi^{\left(\frac{1}{2},0\right)}(\p) \;+ \;\rb {\mathcal A}\; \rb \;
\chi^{\left(0,\frac{1}{2}\right)}(\p) = 0\,;
\eeq
while similar use of Eq. 
(\ref{eq79}) in Eq. (\ref{eq4}) results in:
\beq
\lb\;{\mathcal A}^{-1} \;\lb \; \chi^{\left(\frac{1}{2},0\right)}(\p)
\;-\; \chi^{\left(0,\frac{1}{2}\right)}(\p)  =0\,.
\eeq
The last two equations when combined into a matrix form result in the
{\em momentum-space master equation for\/} $\xi(\p)$,

\beq
\left(
\begin{array}{cc}
- \mathbb{I} & \rb {\mathcal A}\; \rb \\
\lb\;{\mathcal A}^{-1} \;\lb &  - \mathbb{I}
\end{array}
\right)\xi(\p)=0\,.\label{meq}
\eeq
Thus, the momentum-space equation for 
$\xi(\p)$ is entirely determined by the boosts
$\rb$ and $\lb$ and the CPT-property encoding matrix $\mathcal A$. 
Inserting 
$
{\mathcal A}
$
from 
Eq. (\ref{amat}) into (\ref{meq}), 
we evaluate the determinant of the operator 
\beq
{\mathcal O}= \left(
\begin{array}{cc}
- \mathbb{I} & \rb {\mathcal A}\; \rb \\
\lb\;{\mathcal A}^{-1} \;\lb &  - \mathbb{I}
\end{array}
\right)\,,
\eeq
and find it to be:
\beq
Det[{\mathcal O}]= \frac{
\left(
m^2+p^2-(2 m +E)^2
\right)^2\;
\left(m^2+p^2-E^2
\right)^2}{
\left({2 m (E + m)}\right)^4 }
\,,\nonumber\\
\eeq
where $p=\vert\p\vert$.
The wave operator, 
${\mathcal O}$, supports two type of spinors. Those associated
with the usual  dispersion relation, 
\beq
E^2=m^2+p^2\,, \quad \mbox{multiplicity =  4} \label{ds1}
\eeq
and those
associated with:
\beq
E=\cases{-\;2 m -\sqrt{m^2+p^2}\,,\quad \mbox{multiplicity =  2}\cr
	 -\;2 m +\sqrt{m^2+p^2}\,,\quad \mbox{multiplicity  = 2}}\,.\label{nd}
\eeq
The origin of the new dispersion relation must certainly lie, or at least 
we suspect it to be so, in
the new $U(2)$ phases matrix. When ${\mathcal A_\pm}$ equals 
$\pm \mathbb{I}$, i.e.
when we confine to spin-$1/2$ charged particles, 
only the usual dispersion relation gets invoked.
For the extended set of Majorana spinors the situation is more subtle as 
we shall soon discuss. 

There are  also data-dictated reasons which suggest 
a possible violation of Lorentz symmetry, and appearance 
of new dispersion relations. See, e.g.,  the work of 
Mavromatos and Lehnert \cite{nm2002,rl2003}. A recent
review on  theories 
with varying speed of light is by Magueijo \cite{jm2003}.
Though some interpretational question still
remain to be resolved\cite{c1,c2,c3}, the  
work of  Amelino-Camelia on the subject\cite{gac2002} when
extended to Majorana particles shall also carry a 
preferred frame and result in a similar deformation as
noted above. We also draw our reader's attention to systematic 
work of
Mattingly, Jacobson, and Liberati
on violation of Lorentz symmetry and dispersion relations \cite{MJL2002}.
The basic thread running through all these works is that in one
way or antother theory and observations suggest a modification
of $E=\pm \sqrt{m^2+p^2}$. In our work such a modification appears
without invoking  nonlinear realizations of the spacetime symmetries,
and without introducing any additional assumptions.  

\subsection{Dirac equation}

To give confidence to our reader in the physical content of the
Master equation we now apply it to the charged particle 
spinors of Dirac formalism. Once we do that we shall return
to the task of constructing momentum-space wave equation
for the $\lambda(\p)$.

The $\mathcal A$ can be read off from the Dirac rest spinors.
However, we remind the reader, that the writing down of the 
Dirac rest spinors, as shown by Weinberg
and also by our independent studies,
follows from the following two requirements:
(a)
The conservation of parity \cite{SW1995,review,AK2001}; and that
(b)
in a quantum field theoretic framework, the Dirac field describe 
fermions \cite{SW1995}.
These {\em physical requirements\/} determine   
$\mathcal A$ to be:
\beq
\vspace{-\abovedisplayskip}
{\mathcal A} =\cases{+ \;
\mathbb{I}\,,\quad\mbox{for}\,\,u(\p)\,\,\mbox{spinors} \cr
-\;\mathbb{I}\,,\quad\mbox{for}\,\,v(\p)\,\,\mbox{spinors}\,,}
\eeq
and correspond to ${\mathcal A}_+$ with $a=1$, $\phi_a=0$, and
$a=1$, $\phi_a=\pi$, respectively, with $\phi_b$ remaining arbitrary.
The subscript on $\mathcal A$ simply represents that its determinant is 
plus unity.
Using this information in the Master equation (\ref{meq}),
along with the explicit expressions for $\rb$ and $\lb$,
yields:
\beq
\left(
\begin{array}{cc}
-\;\mathbb{I} & \exp\left(\s\cdot\bv\right)\\
\exp\left(-\;\s\cdot\bv\right) & -\;\mathbb{I}
\end{array}
\right)\,u(\p)=0\,,&&\label{ueq} \\
\left(
\begin{array}{cc}
\mathbb{I} & \exp\left(\s\cdot\bv\right)\\
\exp\left(-\;\s\cdot\bv\right) & \mathbb{I}
\end{array}
\right)\,v(\p)=0\,.&&\label{veq}
\eeq
Exploiting the fact that $\s^2=\mathbb{I}$, and using the 
definition of the boost parameter $\bv$  given in Eqs. (\ref{bp}),
the exponentials that appear in the above equation take the form,
\beq
\exp\left(\pm \;\s\cdot\bv\right)
={\left(E\mathbb{I} \pm \s\cdot\p\right)\over m}\,.\label{lin}
\eeq
Using these expansions in Eqs. (\ref{ueq}) and  
(\ref{veq}), multiplying both sides of the resulting equations
by $m$, using $p_\mu=\left(E,-\p\right)$, and
introducing $\gamma^\mu$ as in Eqs. (\ref{gammamatrices}),
gives Eqs. (\ref{ueq}) and (\ref{ueq}) the form 
\beq
\left(p_\mu\gamma^\mu -m \mathbb{I}\right)\,u(\p)=0\,, && \\
\left(p_\mu\gamma^\mu +m \mathbb{I}\right)\,v(\p)=0\,.
\eeq
These are the well-known momentum space wave equations for 
the charged particle spinors (i.e. the Dirac equations). 
The {\em linearity\/} of these equations in $p_\mu$ is due to form of 
$\mathcal A$, and the property of Pauli matrices, $\s^2=\mathbb{I}$  
-- see, Eq.
(\ref{lin}).
The 
$\mbox{Det}\left[p_\mu\gamma^\mu -m \mathbb{I}\right]= 0$
yields dispersion relation (\ref{ds1}) only.

\subsection{Wave equation for the extended set of Majorana spinors}

The requirement that the $\lambda(\p)$ be eigenstates of the 
charge conjugation operator completely determines  $\mathcal A$
for the neutral particle spinors to be:\footnote{This is slightly non trivial
but can be extracted from explicit forms of $\lambda(\0)$ given
in Eqs. (\ref{ls}) and (\ref{la}) and by making use
of the information given in Appendix B.}
\beq
{\mathcal A} = \zeta_\lambda\,\Theta\,\beta\,,
\eeq
where 
\beq
\beta= \left(
\begin{array}{cc}
\exp\left(i\phi\right) & 0 \\
0 & \exp\left(-\;i\phi\right)
\end{array}
\right)\,.
\eeq
Explicitly, 
\beq
{\mathcal A}^S_{-}  =
\left(
\begin{array}{cc}
0 & - i e^{-i \phi} \\
i e^{i\phi} & 0
\end{array}
\right)\,,\quad
{\mathcal A}^A_{-} =
\left(
\begin{array}{cc}
0 &  i e^{-i \phi} \\
- i e^{i\phi} & 0
\end{array}
\right)\,.
\eeq
The noted $\mathcal A$'s correspond to the following choice of the
parameters $\{a,\,\phi_a,\,\phi_b\}$: $a=0,\, \phi_b= -\phi+\pi$ and
$a=0,\, \phi_b= -\phi$, respectively,  with $\phi_a$ remaining arbitrary.
The subscript on $\mathcal A$  is to remind that its determinant is 
minus unity.  
This difference -- summarized in Table 2 --
in $\mathcal A$, for Dirac and Majorana spinors, 
does not allow the 
$\lambda(\p)$ to satisfy the  Dirac equation.
Following the same procedure as in Sec. 5.1 , and using
\beq
\exp\left(\pm\;\frac{\s\cdot\bv}{2}\right)
=
\frac{
\left(E+m\right)\mathbb{I}
\pm \s\cdot\p}
{\sqrt{2m\left(E+m\right)}}\,,
\eeq
we obtain, instead:
\def\op{\left(p_\mu\gamma^\mu + m \gamma^0\right)}
\beq
\left[\op 
\widetilde{\mathcal A} \op - 2m \left(E+m\right)\mathbb{I}\right]\lambda(\p)=0
\,;\label{eqnn}
\eeq
where
\beq
\widetilde{\mathcal A}=\left(
\begin{array}{cc}
\mathbb{O} & {\mathcal A} \\
{\mathcal A}^{-1} & \mathbb{O}
\end{array}
\right)\,.\label{eqn}
\eeq
As a check we verify that 
$
\mbox{Det} \left[\op 
\widetilde{\mathcal A} \op - 2m \left(E+m\right)\mathbb{I}\right] =0$
yields 
dispersion relation  (\ref{ds1}) and also  (\ref{nd}).

\TABLE{
{\begin{tabular}{|c|c|c|c|c|} \hline\hline
{\sc Spinor type}  & $\mbox{Det}[{\mathcal A}]$  
& $a$ & $\phi_a$ & $\phi_b$ \\ \hline\hline
$\mbox{Dirac}$
 $u(\p)$ & $+1$ & $1$ & $0$      & $\mbox{arbitrary}$ \\\hline  
$\mbox{Dirac}$ $v(\p)$ & $+1$ &$1$ & $\pi$      & $\mbox{arbitrary}$ 
\\  \hline
$\mbox{Majorana}$
$\lambda^S(\p)$ & $-1$ & 0 & $\mbox{arbitrary}$ & $-\phi + 
\pi $\\\hline
$\mbox{Majorana}$ $\lambda^A(\p)$ & $-1$ & 0 & $\mbox{arbitrary}$ 
& $-\phi  $
\\
\hline \hline
\end{tabular}}
\label{tab2}
\caption{The parameters $\{a,\,\phi_a,\,\phi_b\}$. See text.}
}

\section{Physical Interpretation }

The existence of (\ref{nd}), and lack of manifest covariance of
Eq. (\ref{eqnn}), seem to be deeply connected.
Following the stated spirit of this paper outlined in 
observation marked ${\cal{O}}_3$ of Sec. 2, we refrain 
from simply overlooking the 
situation.\footnote{I thank 
Abhay Ashtekar and Naresh Dadhich for discussions at IUCAA. In part, 
the content
of this sections reflects that discussion.}
Here we offer what appears to be the 
most natural physical interpretation. 
To put forward our interpretation, 
we define two wave operators, 
\beq
&& {\cal O}_S = \left[\op 
 \widetilde{\mathcal A} \op - 2m \left(E+m\right)\mathbb{I}\right]
_{{\cal A}={\cal A}^S_-} \,,\\
&&{\cal O}_A=
\left[\op 
 \widetilde{\mathcal A} \op - 2m \left(E+m\right)\mathbb{I}\right]
_{{\cal A}={\cal A}^A_-} \,.
\eeq
By construction, 
\beq
{\cal O}_S \lambda^S(\p)=0\,, \quad  {\cal O}_A \lambda^A(\p)=0\,.
\eeq
for $E=\pm\sqrt{p^2 +m^2}$, and assuming $E= \pm\sqrt{p^2 +m^2}$ 
we verify that,
\beq
{\cal O}_S \lambda^A(\p)\ne 0\,, \quad  {\cal O}_S \lambda^A(\p)\ne 0\,.
\eeq
However, if we assert $E= -2 m \pm \sqrt{p^2+m^2}$, 
a brute force calculation shows the result:
\beq
{\cal O}_S \lambda^A(\p)= 0\,, \quad  {\cal O}_S \lambda^A(\p)= 0\,.
\eeq
This role reversal of self and anti-self conjugate sectors
runs miraculously through the whole structure. For instance,
imposing  $E= -2 m \pm \sqrt{p^2+m^2}$ results in (cf., results
of Sec. 4, paying careful attention to the  sub- and super- 
scripts $S$ and $A$):

\beq
\stackrel{\neg}\lambda^A_{\alpha}(\p)\;
\lambda^A_{\alpha^\prime}(\p) = +\; 2 m \;\delta_{\alpha\alpha^\prime}\,,\\
\stackrel{\neg}\lambda^S_{\alpha}(\p)\;
\lambda^S_{\alpha^\prime}(\p) = -\; 2 m\; \delta_{\alpha\alpha^\prime
}\,.
\eeq
The completeness relation also formally 
``interchanges'' self and anti-self conjugate
sectors,
\beq
\frac{1}{2 m}\sum_\alpha 
 \left[\lambda^A_{\alpha}(\p) \stackrel{\neg}\lambda^A_{\alpha}(\p) 
      - \lambda^S_{\alpha} (\p)\stackrel{\neg}\lambda^S_{\alpha}(\p)\right]
 = \mathbb{I}\,, 
\eeq

Our interpretation, then, is the following conjecture: Majorana particles,
in order to minimize energy of the system, physically realize the new 
dispersion relation. The massless limit of the $\lambda^S(\p)$ 
and $\lambda^S(\p)$, when coupled with results of Appendix G, suggest
that we identify self conjugate sector with Majorana particles, while
the anti-self conjugate sector describes Majorana antiparticles. The 
observed cosmological matter-antimatter asymmetry then may be a result
of $S$-$A$ asymmetry contained in the new dispersion relation. 
The lack of manifest covariance of the wave equation for the
extended set of Majorana spinors reflects, and defines, a 
preferred frame which we identify with that of cosmic neutrino
background. It shall then be an experimental challenge to decipher if
the frame of cosmic microwave background coincides with   
that of  cosmic neutrino
background.

The fact that  the new dispersion relation may have escaped experimental
detection may simply be related to the fact direct experiments with
neutrinos are notoriously difficult and that neutrino masses 
\cite{HM2001,HM2002} are
orders of magnitude smaller than their charged counterparts.

\section{Additional properties of the extended 
set of Majorana spin\-ors}

In Majorana realization (``representation''), 
Appendix E shows that the self conjugate $\lambda^S(\p)$
are  real, while antiself conjugate $\lambda^A(\p)$ are pure
imaginary. 
The commutativity of C and P for the extended 
set of Majorana spinors is established in Appenix G, and 
agrees with the old results of Foldy and Nigam \cite{NF1956}.
These results allow to contruct top right block of Table 1.
The parity properties of the extended 
set of Majorana spinors is given in Appendix H. 
We have followed this format so as not to impede the general
flow of arguments and placing some of the results in the 
Appendices is by no means intended to diminish their 
relative significance.

\section{Conclusion}

The standard Dirac field is written as,
\beq
\psi^{charged}(x) = \int \frac{d^3 p}{{2\pi}^3}\frac{m}{p_0}
\sum_{\sigma={+,-}} \left[
a_\sigma(\p) u_\sigma(\p) e^{-p_\mu x^\mu}
+ b_\sigma^\dagger(\p)  v_\sigma(\p) e^{+p_\mu x^\mu}\right]\,.
\eeq
By identifying $b_\sigma^\dagger(\p)$ with $a_\sigma^\dagger(\p)$
one obtains the Majorana field
\beq
\psi^{neutral}(x) = \int \frac{d^3 p}{{2\pi}^3}\frac{m}{p_0}
\sum_{\sigma={+,-}} \left[
a_\sigma(\p) u_\sigma(\p) e^{-p_\mu x^\mu}
+ a_\sigma^\dagger(\p)  v_\sigma(\p) e^{+p_\mu x^\mu}\right]\,.
\eeq
Such a description makes
Majorana particles subordinate to Dirac's representation space in which
the particle and antiparticle spinors are the basis spinors and 
endow the space with very  specific
C, P, and T properties. Inspired by recent 
Heidelberg-Moscow results \cite{HM2001,HM2002}, 
we have aspired to fufill Majorana's original goal by
bringing full symmetry between the charged and 
fundamentally neutral particles. We constructed 
a complete set of Majorana spinors and unearthed their
properties. This allows for introduction of,
\beq
\nu^{neutral}(x) = \int \frac{d^3 p}{{2\pi}^3}\frac{m}{p_0}
\sum_{\alpha=\{-,+\},\{+,-\}} \left[
c_\alpha(\p) \lambda^S_\alpha(\p) e^{-p_\mu x^\mu}
+ c_\alpha^\dagger(\p)  \lambda^A_\alpha(\p) e^{+p_\mu x^\mu}\right]\,,
\eeq
and 
\beq
\nu^{charged}(x) = \int \frac{d^3 p}{{2\pi}^3}\frac{m}{p_0}
\sum_{\alpha=\{-,+\},\{+,-\}} \left[
c_\alpha(\p) \lambda^S_\alpha(\p) e^{-p_\mu x^\mu}
+ d_\alpha^\dagger(\p)  \lambda^A_\alpha(\p) e^{+p_\mu x^\mu}\right]\,.
\eeq
Describing charged particles by 
$\nu^{charged}(x)$ may appear aburd. But it is this ``absurdity''
of $\psi^{neutral}(x)$ that led us to this paper.
In fact neither of the four fields defined above are absurd at any
level. Deciphering thier physical content and realization is 
a matter of further theoretical work, and prediction of
phenomenologically distinct phenomena may open up a new
experimental arena.  It is already clear
that such a venture may be full of unexpected surprises.
We have discovered some surprises already  and  have
 summarized the same in the Abstract
and argued in the paper.

\acknowledgments

Parts of this paper are based    on  Concluding Remarks 
and an Invited Talk presented at ``Physics beyond 
the Standard Model: Beyond the Desert 2002'' and archived as Ref. 
\cite{A2002}.

\appendix{\underline{Appendix A:} Derivation of Eq. (3.9)}

Complex conjugating Eq. (\ref{x})
gives, 
\beq
{\s}^\ast\cdot{{\hp}} \;\left[\phi_L^\pm (\0)\right]^\ast= 
\pm\;\left[\phi_L^\pm(\0)\right]^\ast\,.  
\eeq
Substituting for $\s^\ast$ from Eq. (\ref{wigner}) then results in,
\beq
\Theta \s \Theta^{-1}\cdot{\hp} \,\left[\phi_L^\pm (\0)\right]^\ast
= 
\mp\,\left[\phi_L^\pm(\0)\right]^\ast\, .
\eeq
But $\Theta^{-1} = -\Theta$. So, 
\beq
- \Theta \s \Theta
\cdot{\hp} \,\left[\phi_L^\pm (\0)\right]^\ast= 
\mp\,\left[\phi_L^\pm(\0)\right]^\ast \,.
\eeq
Or, equivalently,
\beq
\Theta^{-1} \s \Theta
\cdot{\hp} \,\left[\phi_L^\pm (\0)\right]^\ast= 
\mp\,\left[\phi_L^\pm(\0)\right]^\ast \,.
\eeq
Finally, left multiplying  both sides of the preceding 
equation by $\Theta$, and moving $\Theta$ through 
${\hp}$, yields Eq. (\ref{y}).

\appendix{\underline{Appendix B:} The $\phi_L^\pm(\0)$}

Representing the unit vector along $\p$, 
as, 
\beq
\hp =\Big(\sin(\theta)\cos(\phi),\,
\sin(\theta)\sin(\phi),\,\cos(\theta)\Big)\,,
\eeq
the $\phi^\pm_L(\0)$ take the explicit form:
\beq
\phi_L^+(\0) =
\sqrt{m} e^{i\vartheta_1} 
\left(
\begin{array}{c}
\cos(\theta/2) e^{-i\phi/2}\\
\sin(\theta/2) e^{i\phi/2}
\end{array}
\right)\,,\\
\phi_L^-(\0) =
\sqrt{m} e^{i\vartheta_2} 
\left(
\begin{array}{c}
\sin(\theta/2) e^{-i\phi/2}\\
-\cos(\theta/2) e^{i\phi/2}
\end{array}
\right)\,.
\eeq
In this paper we take $\vartheta_1$ and $\vartheta_2$ to be zero.

\appendix{{\underline{Appendix C:} Bi-orthonormality 
relations for $\lambda(\p)$ spinors}

On setting $\vartheta_1$ and $\vartheta_2$ to be zero --- a fact
that we explicitly note \cite{AGJ1994,DVA1996} --- we find the following {\em 
bi-orthonormality\/} relations for the self-conjugate 
spinors,
\beq
 \overline{\lambda}^S_{\{-,+\}}(\p) \lambda^S_{\{-,+\}} (\p) = 0\,,
\quad 
\overline{\lambda}^S_{\{-,+\}}(\p) \lambda^S_{\{+,-\}} (\p) = + 2 i m
\,,&& \label{bo1}\\
 \overline{\lambda}^S_{\{+,-\}}(\p) \lambda^S_{\{-,+\}} (\p) = - 2 i m\,,
\quad 
\overline{\lambda}^S_{\{+,-\}}(\p) \lambda^S_{\{+,-\}} (\p) = 0
\,.\label{bo2}&&
\eeq
Their counterpart for antiself-conjugate spinors reads, 
\beq
 \overline{\lambda}^A_{\{-,+\}}(\p) \lambda^A_{\{-,+\}} (\p) = 0\,,
\quad 
\overline{\lambda}^A_{\{-,+\}}(\p) \lambda^A_{\{+,-\}} (\p) = - 2 i m
\,,&&\label{bo3}\\
 \overline{\lambda}^A_{\{+,-\}}(\p) \lambda^A_{\{-,+\}} (\p) = + 2 i m\,,
\quad 
\overline{\lambda}^A_{\{+,-\}}(\p) \lambda^A_{\{+,-\}} (\p) = 0
\,,\label{bo4}&&
\eeq
while all combinations  of the type 
$\overline{\lambda}^A(\p) \lambda^S(\p)$ and
$\overline{\lambda}^S(\p) \lambda^A(\p)$ identically vanish.
We take note that the bi-orthogonal norms of the Majorana spinors
are intrinsically {\em imaginary.}
The associated completeness relation is:
\def\da{{\{+,-\}}}
\def\ua{{\{-,+\}}}

\beq
-\frac{1}{2 i m}&&
{\Bigg(}\left[\lambda^S_{\{-,+\}}(\p) \overline{\lambda}^S_\da(\p)
-\lambda^S_\da(\p) \overline{\lambda}^S_{\{-,+\}}(p)\right]   \nonumber\\ 
&& \quad -
\left[\lambda^A_{\{-,+\}}(\p) \overline{\lambda}^A_\da(\p)
-\lambda^A_\da (\p)\overline{\lambda}^A_{\{-,+\}}(\p)\right]{\Bigg)}
 = \mathbb{I}\,.\nonumber \\ \label{lc}
\eeq

\appendix{\underline{Appendix D:} The $\rho(\p)$ 
spinors}

Now, $(1/2,0)\oplus(0,1/2)$ is a four dimensional representation space.
Therefore, there cannot be more than four independent spinors.
Consistent with this observation, we find that 
the $\rho(\p)$ spinors are related to the $\lambda(\p)$ spinors 
through the following identities:
\beq
&&\rho^S_\ua(\p) = - i \lambda^A_\da(\p)\,,\quad
\rho^S_\da(\p) = + i \lambda^A_\ua(\p),\label{id1}\\
&&\rho^A_\ua(\p) = + i \lambda^S_\da(\p)\,,\quad
\rho^A_\da(\p) = - i \lambda^S_\ua(\p)\,.\label{id2}
\eeq
Using these identities, one may immediately obtain the bi-orthonormality
and completeness relations for the $\rho(\p)$ spinors. 
In the massless limit, $\rho^S_\da(\p)$ and $\rho^A_\da(\p)$
{\em identically vanish.\/}
A particularly simple orthonormality, as opposed to bi-orthonormality,
relation exists between the 
$\lambda(\p)$ and $\rho(\p)$ spinors:
\beq
\overline{\lambda}^S_\ua(\p) 
\rho^A_\ua(\p) = -2 m = \overline{\lambda}^A_\ua(\p)
\rho^S_\ua (\p)\\  
\overline{\lambda}^S_\da(\p) 
\rho^A_\da(\p) = - 2 m = \overline{\lambda}^A_\da(\p)
\rho^S_\da (\p).
\eeq
An associated completeness relation also exists, and it reads:
\beq
-\frac{1}{2  m}&&
{\Bigg(}\left[\lambda^S_\ua(\p) \overline{\rho}^A_\ua(\p)
+\lambda^S_\da(\p) \overline{\rho}^A_\da(p)\right] \nonumber \\
&& \quad+
\left[\lambda^A_\ua(\p) \overline{\rho}^S_\ua(\p)
+\lambda^A_\da (\p)\overline{\rho}^S_\da(\p)\right]{\Bigg)} 
= \mathbb{I}\,.\nonumber\\
\label{lrcompleteness}
\eeq
The results of this section are in spirit of Refs.~
\cite{Mc1957,Case1957,AGJ1994,DVA1996}.

The completeness relation (\ref{lc}) confirms
that a physically complete theory of 
fundamentally neutral particle spinors must incorporate the 
self as well as antiself conjugate spinors. However, one has a choice.
One may either work with the set $\{\lambda^S(\p),\lambda^A(\p\})$,
or with the physically and mathematically equivalent set,
  $\{\rho^S(\p),\rho^A(\p)\}$. One is also free to choose some 
appropriate combinations of neutral particle spinors 
from these two sets.

\appendix{\underline{Appendix E:} Extended set of
Majorana spinors in Majorana realization}

The $\lambda^{S,A}(\p)$ obtained above are in Weyl realization  
(subscripted by, $W$.
In  Majorana realization (subscripted by, $M$) these spinors are given by:
\beq
\lambda^{S,A}_{M}(\p) = {\mathcal S} \,\lambda^{S,A}_{W}(\p)\,,
\eeq
where
\beq
{\mathcal S} = \frac{1}{2} \left(
\begin{array}{cc}
\mathbb{I} + i\Theta &{\;\;} \mathbb{I} - i\Theta \\
-\left(\mathbb{I} - i\Theta\right) & {\;\;}\mathbb{I} + i\Theta 
\end{array}\right)\,.
\eeq 
Calculations show
$\lambda^S_M(\p)$ are real, while $\lambda^{A}_{M}(\p)$
are pure imaginary.

\appendix{\underline{Appendix F:}  The $\lambda(\p)$ do
not satisfy Dirac equation}
\footnote{The main result of this Appendix is a re-rendering of a proof given
my M. Kirchbach\cite{mkpc}. Any mistake, if any, that the reader may 
notice is entirely due to our failure.}

The bi-orthonormality relations (\ref{bo1}-\ref{bo4})
and the completeness relation (\ref{lc}) are counterpart
of the following relations for the charged, i.e. Dirac, particle spinors:
\beq
&& \overline{u}_h(\p)\, u_{h^\prime}(\p) =  +\;2 m\;
 \delta_{h {h^\prime}}\,,\label{d1}\\
&&\overline{v}_h(\p)\, v_{h^\prime}(\p) =  -\;2 m\;
 \delta_{h {h^\prime}}\,,\label{5}\\
&& \frac{1}{2 m}
\left[\sum_{h=\pm 1/2}
 u_{h}(\p) \overline{u}_h(\p) -
 \sum_{h=\pm 1/2} v_{h}(\p) \overline{v}_h(\p)\right] = \mathbb{I}\,.\label{1}
\eeq
Furthermore, if one wishes (with certain element of hazard to become 
apparent below), one can 
write the the {\em momentum-space} extended set of Majorana spinors
 $\{\lambda^S(\p),\lambda^A(\p\})$, in terms of Dirac spinors in
{\em momentum-space}, $\{u(\p),v(\p)\}$. This task is best
accomplished by introducing the following notation:
\beq
&& d_1\equiv u_+(\p), \,
d_2\equiv u_-(\p), \,
d_3\equiv v_+(\p), \,
d_4\equiv v_-(\p)\,, \\
&& m_1\equiv \lambda^S_\ua(\p), \,
m_2\equiv \lambda^S_\da(\p), \,
m_3\equiv \lambda^A_\ua(\p), \,
m_4\equiv \lambda^A_\da(\p)\,.
\eeq
Then, the extended set of Majorana  spinors can be written as,
\beq
m_i= \sum_{j=1}^4 \Omega_{ij} d_j\,,\quad i=1,2,3,4\label{md}
\eeq
where

\begin{equation}
\Omega_{ij}=
 \cases{
+\left({1}/{2 m}\right) \overline{d}_j\, m_i \mathbb{I}\,, & \quad{\mbox for}
\,\, $j =1,2$\cr
-\left({1}/{2 m}\right) \overline{d}_j \,m_i \mathbb{I}\,, & \quad{\mbox for} 
\,\,$j =3,4$\cr}\,.
\end{equation}
In matrix form,  the $\Omega$ reads:
\beq
\Omega =
\frac{1}{2}\left(
\begin{array}{cccc}
\mathbb{I} & - i\mathbb{I} & - \mathbb{I} & - i\mathbb{I} \\
i\mathbb{I} & \mathbb{I} & i\mathbb{I} & - \mathbb{I} \\
\mathbb{I} & i\mathbb{I} & - \mathbb{I} & i\mathbb{I} \\
- i\mathbb{I} & \mathbb{I} & - i\mathbb{I} & - \mathbb{I}
\end{array}
\right)\,.\label{omega}
\eeq
Equations (\ref{md}) and (\ref{omega}) immediately tell us that {\em a
spinor\/} in the extended set of Majorana spinors is a linear combination
of the  Dirac {\em particle and antiparticle\/} 
spinors. In momentum space, the Dirac spinors 
are annihilated by $\left(\gamma^\mu p_\mu \pm m \mathbb{I}\right)$,
\beq
\cases{
\mbox{For particles:}
\quad\left(\gamma^\mu p_\mu - m \mathbb{I}\right) u(\p)=0\,,\quad\cr
\mbox{For antiparticles:}\quad
\left(\gamma^\mu p_\mu + m \mathbb{I}\right) v(\p)=0\,.\cr}\label{deqs}
\eeq
Since the mass terms carry opposite signs, 
hence are different for the particle and antiparticle, 
the spinors in the extended set of Majorana spinors 
cannot be annihilated by 
$\left(\gamma^\mu p_\mu - m \mathbb{I}\right)$, or, by
$\left(\gamma^\mu p_\mu + m \mathbb{I}\right)$. Moreover,
in the configuration space,
since the time evolution of the of $u(\p)$ occurs via
$\exp(-ip_\mu x^\mu)$ while that for   
$v(\p)$ spinors occurs via $\exp(+ip_\mu x^\mu)$ one cannot naively go from
momentum-space expression (\ref{md}) to its configuration space counterpart.
In fact several conceptual and technically subtle hazards 
are confronted if one begins to mix the two set of spinors. One ought to
develop the theory of fundamentally neutral particle spinors
entirely in its own right. We thus end this digression by making part
of the above argument more explicit. For that purpose we introduce:
\beq
M\equiv
\left(
\begin{array}{c}
m_1\\
m_2\\
m_3\\
m_4
\end{array}
\right)\,,\quad
D\equiv
\left(
\begin{array}{c}
d_1\\
d_2\\
d_3\\
d_4
\end{array}
\right)\,,
\Lambda\equiv
\left(
\begin{array}{cccc}
\gamma_\mu p^\mu & \mathbb{O} & \mathbb{O} & \mathbb{O} \\
 \mathbb{O} & \gamma_\mu p^\mu & \mathbb{O} & \mathbb{O}  \\
  \mathbb{O} & \mathbb{O}& \gamma_\mu p^\mu & \mathbb{O}   \\
  \mathbb{O}& \mathbb{O} & \mathbb{O} & \gamma_\mu p^\mu   \\
\end{array}
\right)
\,.
\eeq
In this language, equation (\ref{md}) becomes
\beq
M=\Omega D\,.
\eeq
Now, applying from left the operator $\Lambda$ and using,
$\left[\Lambda,\Omega\right] =0$, we get
\beq
\Lambda M = \Omega \Lambda D\,.
\eeq
But, Eqs. (\ref{deqs}) imply
\beq
\Lambda D = 
\left(
\begin{array}{cccc}
 m\mathbb{I} & \mathbb{O} & \mathbb{O} & \mathbb{O} \\
 \mathbb{O} &  m\mathbb{I} & \mathbb{O} & \mathbb{O}  \\
  \mathbb{O} & \mathbb{O} & - m \mathbb{I} & \mathbb{O}   \\
  \mathbb{O}& \mathbb{O} & \mathbb{O} &    - m \mathbb{I} \\
\end{array}
\right)\,D\,.\label{above}
\eeq
Therefore, on using $D=\Omega^{-1} M$ we obtain,
\beq
\Lambda M = \Omega \left(\mbox{r.h.s. of Eq. \ref{above}}\right) \Omega^{-1} M
\,.
\eeq
An explicit evaluation of, 
$\mu\equiv \Omega \left(\mbox{r.h.s. of Eq. \ref{above}}\right) \Omega^{-1}$,
reveals it to be,
\beq
\mu= \left(
\begin{array}{cccc}
 \mathbb{O}& -i m \mathbb{I} & \mathbb{O} & \mathbb{O} \\
 i m \mathbb{I} &  \mathbb{O} & \mathbb{O} & \mathbb{O}  \\
  \mathbb{O} & \mathbb{O}& \mathbb{O} & i m \mathbb{I}   \\
  \mathbb{O}& \mathbb{O} & -i m \mathbb{I} &    \mathbb{O} \\
\end{array}
\right)\,.
\eeq
Thus, finally giving us the result,
\beq
\left(
\begin{array}{cccc}
\gamma_\mu p^\mu & \mathbb{O} & \mathbb{O} & \mathbb{O} \\
 \mathbb{O} & \gamma_\mu p^\mu & \mathbb{O} & \mathbb{O}  \\
  \mathbb{O} & \mathbb{O}& \gamma_\mu p^\mu & \mathbb{O}   \\
  \mathbb{O} & \mathbb{O} & \mathbb{O} & \gamma_\mu p^\mu   \\
\end{array}
\right)
\left(
\begin{array}{c}
\lambda^S_\ua(\p) \\
\lambda^S_\da(\p) \\
\lambda^A_\ua(\p) \\
\lambda^A_\da(\p)
\end{array}
\right)
-
im \left(
\begin{array}{c}
- \,\lambda^S_\da(\p) \\
 \,\lambda^S_\ua(\p) \\
 \,\lambda^A_\da(\p) \\
- \,\lambda^A_\ua(\p)
\end{array}
\right) = 0\,,\nonumber \\
\eeq
which explicitly establishes the result that 
$\left(\gamma^\mu p_\mu \pm m \mathbb{I}\right)$ do not annihilate
the neutral particle spinors.\footnote{The result contained in the above 
equation confirms earlier result of Ref. \cite{VVD1995}.} 
The text-book assertions that Majorana mass term is 
`off-diagonal'' is a rough translation of this equation.

\appendix{\underline{Appendix G:} 
Commutativity of C and P for the extended set of Majorana 
spinors}

The parity operation is slightly subtle for neutral particle spinors.
In the $(1/2,0)\oplus(0,1/2)$ representation space it reads,
\beq
P= e^{i\phi_P} \gamma^0 {\mathcal R}\,.
\eeq
The ${\mathcal R}$ is defined as,
\beq
{\mathcal R} \equiv
\left\{ \theta\rightarrow\pi-\theta, \;\phi\rightarrow \phi+\pi,\;
p\rightarrow p\right\}\,.
\eeq
This has the consequence that eigenvalues, $h$,  of the helicity
operator
\beq
{\mathbf h} = \frac{\s}{2}\cdot\hp
\eeq
change sign under the operation of $\mathcal R$,
\beq
{\mathcal R}: h \rightarrow h^\prime = - h\,.
\eeq
Furthermore, 
\beq
P u_h(\p) = e^{i\phi_P} \gamma^0 {\mathcal R} u_h(\p) =
 e^{i\phi_P} \gamma^0 u_{-h}(-\p) = 
- i e^{i\phi_P} u_h(\p)
\eeq  
Similarly, 
\beq
P v_h(\p) = i e^{i\phi_P} v_h(\p)\,.
\eeq
We now require the eigenvalues of the $P$ to be real. This
fixes the phase factor,
\beq
e^{i\phi_P} = \pm i\,. \label{pf}
\eeq
The remaining ambiguity, as contained in the sign, 
still remains.  It is  fixed by recourse to text-book 
convention by taking the sign on the right-hand side
of Eq. (\ref{pf}) of to be positive. This very last choice shall
not affect our conclusions (as it should not). 
The parity operator is thus fixed to be,
\beq
P= i \gamma^0 {\mathcal R}\,.
\eeq
Thus, 
\beq
&& P u_h(\p) = +\, u_h(\p)\,,\label{peqa}\\
 && P v_h(\p) = -\, v_h(\p)\,.\label{peqb}
\eeq 
The consistency of  Eqs.  (\ref{peqa}) and (\ref{peqb}) requires,
\beq
\mbox{\sc Charged particle spinors}: \quad P^2= \,\mathbb{I}\,,\quad
[\mbox{{\em cf.} Eq.(\ref{cf2})}]\,.\label{cf1}
\eeq 

To calculate the anticommutator, $\{C,P\}$, when acting on
the $u_h(\p)$ and $v_h(\p)$ we now need, in addition,
the action of $C$ on these spinors. This action
can be summarized as follows:
\beq
C :{\Bigg\{}\begin{array}{l}
  u_{+1/2}(\p) \rightarrow - v_{-1/2}(\p)\,,
 u_{-1/2}(\p) \rightarrow   v_{+1/2}(\p)\,,\\
  v_{+1/2}(\p) \rightarrow   u_{-1/2}(\p) \,,
 v_{-1/2}(\p) \rightarrow - u_{+1/2}(\p)\,.\label{ceq}
\end{array}
\eeq
Using Eqs. (\ref{peqa}), (\ref{peqb}),  
and (\ref{ceq}) one can readily obtain the 
action of anticommutator, $\{C,P\}$, on the four $u(\p)$ and 
$v(p)$ spinors. For each case it is found to vanish: $\{C,P\}=0$.

\bigskip\noindent
The P acting on the  neutral particle spinors
yields the result,
\beq
P\lambda^{S}_\ua (\p)= +\, i\, \lambda^A_\da(\p)\,,
P\lambda^{S}_\da (\p)= -\, i \,\lambda^A_\ua(\p)\,,&&\label{peqla}\\
P\lambda^{A}_\ua (\p)= - \,i \,\lambda^S_\da(\p)\,,
P\lambda^{A}_\da (\p)= + \,i \,\lambda^S_\ua(\p)\,.\label{peqlb}
\eeq
Following the same procedure as before, we now use  (\ref{peqla}),  
(\ref{peqlb}), and (\ref{ceq}) to 
evaluate the action of the commutator $[C,P]$ 
on each of the four neutral particle spinors.
We find it vanishes for each of them: $[C,P] =0$.
It confirms the claim we made in Table \ref{tab1}.

The commutativity and anticommutatitvity of the $C$ and $P$ operators
is a deeply profound result and it establishes that the theory of
neutral and charged particles must be developed in their own rights.
This is the task we have undertaken and are developing here in this
{\em Paper.\/}

\appendix{\underline{Appendix H:} 
\label{pa}Parity asymmetry for the extended set of Majorana 
spinors}

Unlike the charged particle spinors,  Eqs.  (\ref{peqla}) and (\ref{peqlb}) 
reveal that neutral particle spinors are not eigenstates of $P$.   
Furthermore, a rather apparently  
paradoxical asymmetry is contained in these equations.
For instance, the second equation in (\ref{peqla}) reads:
\beq
P\lambda^{S}_\da (\p)= -\, i \,\lambda^A_\ua(\p)\,.
\eeq
Now, in a normalization-independent manner
\beq
\lambda^{S}_\da (\p) \propto 
\left(1+\frac{ \vert \p \vert}{E+m}\right)
\lambda_\da^S(\0)\,,
\eeq
while
\beq
\lambda^A_\ua(\p) \propto
\left(1-\frac{ \vert \p \vert}{E+m}\right)
\lambda_\da^A(\0)\,.
\eeq
Consequently, in the massless/high-energy limit the $P$-reflection of 
$\lambda^{S}_\da (\p)$ identically vanishes. 
The same happens to the $\lambda^{A}_\da (\p)$ spinors
under $P$-reflection. This situation is in sharp
contrast to the charged particle spinors.
The consistency of  Eqs.  (\ref{peqla}) and (\ref{peqlb}) requires 
$P^2 = - \mathbb{I}$ and in the process shows that the remaining two, i.e.
first and third equation in that set, do not contain additional
physical content:
\beq
\mbox{\sc Neutral particle spinors}: \quad P^2=-\,\mathbb{I}\,.
\quad
[\mbox{{\em cf.} Eq.(\ref{cf1})}]\,.\nonumber \\ \label{cf2}
\eeq
That is, for neutral particle spinors:
\beq
\mbox{\sc Neutral particle spinors}: \quad P^4=\,\mathbb{I}\,,\,.\label{cf3}
\eeq

The origin of the asymmetry under $P$-reflection 
resides in the fact that the $(1/2,0)\oplus(1/2,0)$
neutral particles spinors, in being dual helicity objects, 
combine Weyl spinors of {\em opposite\/}
helicities. However,  in the massless limit, the structures of 
$\kappa^{\left(\frac{1}{2},0\right)}$ and 
$\kappa^{\left(0,\frac{1}{2}\right)}$ 
force only positive helicity $\left({1}/{2},0\right)$-Weyl 
and negative helicity $\left(0,{1}/{2}\right)$-Weyl spinors
to be non-vanishing.
For this reason, in 
the massless limit the neutral particle spinors, $\lambda^S_\ua(\p)$
and  $\lambda^A_\ua(\p)$,
carrying negative helicity $\left({1}/{2},0\right)$-Weyl 
and positive helicity $\left(0,{1}/{2}\right)$-Weyl spinors
identically vanish.

So we have the following situation: The $(1/2,0)\oplus(0,1/2)$ 
is a $P$ covariant representation space.  Yet, in the  neutral
particle formalism, it carries
$P$-reflection asymmetry. 
This circumstance has a precedence in the Velo-Zwanziger observation, who
noted \cite{VZ1969},
``the main lesson to be drawn from our analysis is that special relativity
is not automatically satisfied by writing equations which transform 
covariantly.'' We conjecture that this asymmetry may underlie the
phenomenologically known parity violation.
Even though the latter is incorporated, by
hand, in the standard model of the electroweak interactions its true 
physical origin has remained unknown.

\end{document}